\documentclass[a4paper,11pt]{article}
\usepackage{jheppub}
\usepackage{feynmp-auto}
\usepackage[T1]{fontenc}
\usepackage{graphicx}
\usepackage{amsmath}
\usepackage{amssymb}
\usepackage{amsfonts}
\usepackage{bm}
\usepackage{color}

\def\be{\begin{equation}}
\def\ee{\end{equation}}
\def\bea{\begin{eqnarray}}
\def\eea{\end{eqnarray}}

\def\p{\partial}
\newcommand{\mc}[1]{\mathcal{#1}}
\newcommand{\f}[2]{\frac{#1}{#2}}

\title{\boldmath One-loop corrections to vector Galileon theory}

\author[~]{Farid Charmchi${}^\alpha$}
\author[~]{\!\!, Zahra Haghani${}^\beta$}
\author[~]{\!\!, Shahab Shahidi${}^\beta$}
\author[~]{and Leila Shahkarami${}^\beta$}

\affiliation[~]{${}^\alpha$School of Particles and Accelerators, Institute for 
Research in Fundamental Sciences (IPM), P.O. Box 19395-5531, Tehran, Iran}
\affiliation[~]{${}^\beta$School of Physics, Damghan University, Damghan, 
41167-36716, Iran}

\emailAdd{charmchi@ipm.ir}
\emailAdd{z.haghani@du.ac.ir}
\emailAdd{s.shahidi@du.ac.ir}
\emailAdd{l.shahkarami@du.ac.ir}

\abstract{
The effective action of the recently proposed vector Galileon theory is considered. Using the background field method, we obtain the one-loop correction to the propagator of the Proca field from vector Galileon self-interactions. Contrary to the so-called scalar Galileon interactions, the two-point function of the vector field gets renormalized at the one-loop level, indicating that there is no non-renormalization theorem in the vector Galileon theory. Using dimensional regularization, we remove the divergences and obtain the counterterms  of the theory. The finite term is analytically calculated, which modifies the propagator and the mass term and generates some new terms also.
}

\begin{document}
\maketitle
\flushbottom

\section{Introduction}
General relativity (GR) at larger than solar system scales is not perfect. 
Among several deviations of GR from cosmological observations, the 
self-accelerated expansion of the universe has received many attentions. This has been also proved precisely by observations from Planck \cite{planck} and also BICEP2 \cite{bicep}. 
It is now widely accepted that 
this problem can only be explained by introducing 
new degrees of freedom to the theory of gravity. On the other hand, we need another self-accelerating phase at the very early universe which can be achieved by introducing a new scalar degree of freedom, known as inflaton. The late time 
acceleration of the universe can also be explained in a similar way, by adding 
at least one scalar degree of freedom. In this sense, the scalar-tensor (and 
more generally scalar-vector-tensor) theories of gravity can be considered as a 
good starting point to generalize GR at large scales, hence the IR modification of gravity.

In 1961, Brans 
and Dicke developed a gravitational action with a dynamical gravitational constant 
\cite{brans} which gave birth to the scalar-tensor theories. One interesting question is that what kind of interactions of scalar 
field can be added to GR, in order to keep the theory healthy, i.e. free from 
instabilities.  The necessary condition is to write all interactions of a scalar field in flat space-time 
which leads to second order field equations, i.e. at most second order in time 
derivatives. This will avoid Ostrogradski ghost instability at first place \cite{ostro}, which 
has been done in \cite{gali} where the authors prove that there are 5 
independent combinations in 4D satisfying the above condition. The interesting 
fact about these terms is that the scalar field $\pi$ possesses a new symmetry 
$\partial_\mu\pi\rightarrow\partial_\mu\pi+c_\mu$, with $c_\mu$ being some constant. 
This is very similar to Galilean symmetry and hence the scalar field $\pi$ 
dubbed the ``Galileon''. 

Looking at the Galileon interactions in more details, one can see that the first and second terms of the Galileon interactions are the tadpole and the canonical kinetic term. The third term turns out to be the self-interaction term which was previously obtained in the context of DGP brane gravity \cite{dgp}. The fourth and fifth terms are new interaction terms with four and five scalar fields. As mentioned before, the Galileon interactions are written in flat space-time background, and the 
absence of ghost is highly related to the interchange of partial derivatives \cite{gali}. 
This suggests that the minimal substitution of partial derivatives with covariant 
derivatives results in the appearance of higher than second order time derivatives in the 
equations of motion, and hence the ghost comes back \cite{defa}. In order to 
solve this problem, one should add some higher order terms to the Galileon
interactions to make the equations of motion again second order. Hence this will 
breaks the Galilean symmetry of the scalar field 
$\pi$ \cite{defa}. The resulting action is well-known as the ``Horndeski'' action \cite{horn}.
Galileon interactions can be generalized further by coupling the original terms 
to arbitrary functions of $\partial_\mu\pi\partial^\mu\pi$ and $\pi^2$ \cite{gene}. 
Several aspects of the Galileons have been investigated in the literature, consisting 
the cosmological consequences \cite{cosmo}, the black hole solutions 
\cite{blackk} and other aspects of the theory \cite{other}.

Considering the Galileons as a set of possible classical interactions of a scalar 
field, one can ask what happens to the Galileon terms at the quantum level. It is 
well-known that the loop corrections to the Galileon terms do not renormalize 
the galileon coupling, which is  dubbed as a non-renormalization theorem 
\cite{nonrenor}. In \cite{prdplb}, the authors have calculated the one-loop 
effective action of a scalar action with Galileon interactions. It is then 
concluded that the renormalized Fourier space propagator of the scalar field 
will be modified by terms proportional to $k^4$, $k^6$ and $k^8$, confirming that the non-renormalization theorem holds. In 
\cite{heisenmatter}, the one-loop corrections to the 
Galileons by matter loops have been calculated.

It is straightforward to think of interactions for a vector field $A_\mu$ with 
Galileon property. In this sense, we need the most general self-interactions of a 
vector field which result in a second order field equation. These interactions 
can be called ``vector Galileons''. In \cite{nogo}, the authors have proved that 
the assumption of $U(1)$ symmetry for a given vector field restricts the interaction terms to
the canonical Maxwell kinetic term $-1/4 F_{\mu\nu}F^{\mu\nu}$. However, 
relaxing the $U(1)$ symmetry of the action, one can obtain such vector 
Galileon interactions. The 
resulting theory has then acquire 3 degrees of freedom \cite{heisen}. These interaction terms can 
then be generalized by coupling the vector Galileon terms to arbitrary 
functions of $A_\mu A^\mu$. Many works have been done in the literature including the cosmology of vector Galileons \cite{veccos}, its Higgs mechanism \cite{higgs}, its relation to the bi-connection theory \cite{nima} and also the covariantization of the theory \cite{koya}. One of our aims in this paper is to ask which properties of the scalar Galileons in flat space will be inherited to the flat space vector Galileons. Among Galileon symmetry, second order field equations and non-renormalizable theorem, the vector Galileon interactions have only the second property, namely the production of the second order field equation which has been proved in \cite{heisen}. The Galileon symmetry is obviously broken in the theory. The non-renormalization theorem will be investigated in this paper. It seems that the vector Galileon generalization of the original Galileon interactions is not quite faithful.

It is the aim of this paper to calculate the one-loop effective action of a 
vector field theory with vector Galileon self-interactions. We will consider a 
special case where all vector Galileon interactions get multiplied overally to 
$A_\mu A^\mu$. With this choice, we will see that only three out of four 
interaction terms will contribute to the part of effective action containing only two classical fields. Among these terms, 
one is the usual mass term of the vector field and the other two will be its 
self-interactions. We will see that the propagator and the mass term are modified at the one-loop level, which means that the non-renormalization theorem does not hold in the vector Galileon theory.

The next section will be devoted to introducing the model which will be followed by the computation of the one-loop effective action with two classical fields in section \ref{sec3}. We will then summarize our results in section \ref{conclu}.

\section{The model}
We can extend the Proca theory with non-linear ghost-free derivative
interactions of the vector field.
There are only four of them that do not introduce any ghosts to the theory, the ``vector Galileons'', which have the following structures:
\begin{align}\label{vecGalProcaField1}
\mathcal L_2 & = f_2,\nonumber\\
\mathcal L_3 & = f_3 \;\; \partial\cdot \mc{A}, \nonumber\\
\mathcal L_4  &=  f_4\;\left[(\partial\cdot \mc{A})^2+c_1\partial_\rho 
\mc{A}_\sigma \partial^\rho \mc{A}^\sigma-(1+c_1)\partial_\rho \mc{A}_\sigma 
\partial^\sigma \mc{A}^\rho\right],   \nonumber\\
\mathcal L_5  &=f_5\big[(\partial\cdot \mc{A})^3-3c_2(\partial\cdot 
\mc{A})\partial_\rho \mc{A}_\sigma \partial^\rho 
\mc{A}^\sigma-3(1-c_2)(\partial\cdot \mc{A})\partial_\rho \mc{A}_\sigma 
\partial^\sigma \mc{A}^\rho \nonumber\\
&+\left(2-3c_2\right)\partial_\rho \mc{A}_\sigma \partial^\gamma 
\mc{A}^\rho\partial^\sigma \mc{A}_\gamma
+3c_2\partial_\rho \mc{A}_\sigma \partial^\gamma 
\mc{A}^\rho\partial_\gamma \mc{A}^\sigma\big],
\end{align}
where $f_2$ can be any function of $\mc{A}_\mu \mc{A}^\mu$, $F_{\mu\nu}F^{\mu\nu}$, $\epsilon_{\mu\nu\rho\sigma}F^{\mu\nu}F^{\rho\sigma}$ and any possible contraction of $\mc{A}_\mu$ and $F_{\mu\nu}$ where $F_{\mu\nu}=\partial_\mu \mc{A}_{\nu}-\partial_\nu \mc{A}_\mu$. The other functions $f_3,~f_4$ and $f_5$ should depend only on $\mc{A}_\mu \mc{A}^\mu$ \cite{heisen}. In this work we restrict ourselves to the case $f_2=\mc{A}^2$ to make the vector Galileon massive. $f_3$ can not be unity because $\mc{L}_3$ becomes total derivative in this case. In order to keep track of the $\mc{L}_3$ term we will set $f_3=\mc{A}^2$. The choice of $f_4$ and $f_5$ does not change the qualitative results of this paper. As we will see in the following, the $\mc{L}_3$ term itself will modify the propagator of the vector Galileon. We then assume that $f_4,f_5=\mc{A}^2$.
In the following sections, we are going to calculate the one-loop corrections to the propagater of the vector field $\mc{A}_\mu$.
For this purpose we only need to take into account the terms with four powers of $\mc{A}_\mu$ and therefore the Lagrangian $\mc{L}_5$ does not contribute to our analysis.  
By linearly combining the relevant terms, our action is as follows:
\begin{align}\label{2}
S=\int d^4x\left[ -\f14 F_{\mu\nu}F^{\mu\nu}-\f12m^2\mc{A}_\mu 
\mc{A}^\mu+\f12\alpha_3\mc{L}_3+\f12\alpha_4\mc{L}_4\right],
\end{align}
where the first and second terms are the standard kinetic and mass ($\mathcal L_2$) terms for a vector field, respectively. 

\section{Calculation of the one-loop corrections}\label{sec3}
In this section, we use the background field method to arrive at our main purpose, i.e.\;the calculation of the one-loop corrections of the theory given in the previous section.
Consider the following perturbed vector field
\begin{align}\label{3}
\mc{A}_\mu=A_\mu+B_\mu,
\end{align}
in which $B_\mu$ is the fluctuation around the classical background field  $A_\mu$.
Substituting (\ref{3}) into (\ref{2}), expanding the action up to second order in $B_\mu$ and neglecting total derivatives, we obtain the bilinear form of the Lagrangian as
\begin{align}\label{4}
\mc{L}=\f12 B^\mu\bigg[(\Box-m^2)\eta_{\mu\nu}-\partial_\mu\partial_\nu+\alpha_3\mc{L}^{cl}_{3\mu\nu}+\alpha_4\mc{L}^{cl}_{4\mu\nu}\bigg]B^\nu,
\end{align}
where
\begin{align}\label{5}
\mc{L}^{cl}_{3\mu\nu}=2A_\mu\partial_\nu+\partial^\alpha A_\alpha \eta_{\mu\nu},
\end{align}
and
\begin{align}\label{6}
\mc{L}^{cl}_{4\mu\nu}=&-\p_\mu A^2\p_\nu-A^2\p_\mu\p_\nu-c_1\eta_{\mu\nu}\big(\p^\rho A^2\p_\rho+A^2\Box\big)+4A_\mu(\p.A)\p_\nu+4c_1A_\mu\p_\rho A_\nu\p^\rho\nonumber\\&+\eta_{\mu\nu}\bigg[(\p.A)^2+c_1\p_\rho A_\sigma\p^\rho A^\sigma-(1+c_1)\p_\rho A_\sigma\p^\sigma A^\rho\bigg]\nonumber\\&+(1+c_1)\big[\p_\nu A^2\p_\mu+A^2\p_\mu\p_\nu-4A_\mu\p_\nu A_\rho\p^\rho\big].
\end{align}
Therefore, the one-loop effective action would be expressed as
\begin{align}\label{7}
\Gamma_1&=\textmd{tr}\log\bigg[\eta_{\mu\nu}(\Box-m^2)-\p_\mu\p_\nu
+\alpha_3\mc{L}^{cl}_{3\mu\nu}+\alpha_4\mc{L}^{cl}_{4\mu\nu}\bigg]\nonumber\\
&=\textmd{tr}\log\big[\mc{O}_{\mu\nu}\big]+\textmd{tr}\log\bigg[1+\alpha_3\mc{O}^{-1\mu\nu}\mc{L}^{cl}_{3\rho\nu}
+\alpha_4\mc{O}^{-1\mu\nu}\mc{L}^{cl}_{4\rho\nu}\bigg].
\end{align}
where $\mc{O}_{\mu\nu}\equiv\eta_{\mu\nu}(\Box-m^2)-\p_\mu\p_\nu$
and the inverse of this operator can be obtained as
\begin{align}\label{8}
\mc{O}^{-1\mu\nu}=\f{1}{\Box-m^2}\left[\eta^{\mu\nu}-\f{1}{m^2}\partial^\mu\partial^\nu\right].
\end{align}

Our goal is to calculate the divergent part of the one-loop effective action presented in (\ref{7}), in the second order in $A_\mu$.
Notice that the first term in \eqref{7} contributes to the vacuum energy which we will neglect in the following. 
Expanding the Logarithm in the second term of the one-loop effective action and keeping only the quadratic parts will result in
\begin{align}\label{9}
-\f12\alpha_3^2\,\textmd{tr}\left[\mc{O}^{-1\mu\rho}\mc{L}^{cl}_{3\rho\sigma}\mc{O}^{-1\sigma\alpha}
\mc{L}^{cl}_{3\alpha\nu}\right]+
\alpha_4\,\textmd{tr}\left[\mc{O}^{-1\mu\rho}\mc{L}^{cl}_{4\rho\sigma}\right].
\end{align}
In what follows, we calculate the divergent parts of these two terms.
Let us first discuss the first term, which will be called $\mc{X}_3$ in the following.
By Fourier transforming the fields and taking the trace of this term, one can obtain
\begin{align}\label{10}
\mc{X}_3\equiv&-\f12\alpha_3^2\,\textmd{tr}\left[\mc{O}^{-1\mu\rho}\mc{L}^{cl}_{3\rho\sigma}
\mc{O}^{-1\sigma\alpha}\mc{L}^{cl}_{3\alpha\nu}\right]\nonumber\\
=&\,\f12\alpha_3^2\,\int \f{d^4p}{(2\pi)^4} \int d^4k\,\f{1}{p^2+m^2}\left[\eta^{\nu\rho}+\f{p^\nu p^\rho}{m^2}\right]\big[2A_\rho(-k)(p+k)_\alpha-\eta_{\rho\alpha}A_\mu(-k)k^\mu\big]\nonumber\\&\qquad\times\f{1}{(p+k)^2+m^2}\left[\eta^{\alpha\beta}+\f{(p+k)^\alpha (p+k)^\beta}{m^2}\right]\big[2A_\beta(k)p_\nu+\eta_{\beta\nu}A_\lambda(k)k^\lambda\big].
\end{align}
Notice that this integral contains a product of two different quadratic factors in the denominator.
To deal with them, we take advantage of Feynman parameterization technique.
Taking this into account and then shifting the integration variable $p_\mu$ to $p_\mu=l_\mu-xk_\mu$ (where $x$ is the Feynman parameter), one can obtain
\begin{align}\label{11}
\mc{X}_3=\f12\alpha_3^2\int d^4k\int \f{d^dl}{(2\pi)^d}\int_0^1 dx\,\f{1}{[l^2+\Delta]^2}\big(L_0+L_2 l^2+L_4 l^4\big),
\end{align}
where $\Delta=m^2+x(1-x)k^2$. Here $L_i\equiv L_{i}^{\alpha\beta}A_\alpha(k) A_\beta(-k)$ do not contain $l_\mu$ and their full form are reported in the Appendix. 
To calculate the loop integral over $l$, which obviously is divergent, we apply dimensional regularization. To that end we have changed the dimension of the integration from $4$ to $d$.

Performing the integral over $l_\mu$, one can arrive at
\begin{align}\label{12}
\mc{X}_3=\f{i\alpha_3^2}{2}\f{\Gamma\left(2-\f{d}{2}\right)}{(4\pi)^{\f{d}{2}}}\int d^4k\int_0^1 dx\bigg[L_0\left(\f{1}{\Delta}\right)^{2-\f{d}{2}}
+L_2\f{d}{d-2}\left(\f{1}{\Delta}\right)^{1-\f{d}{2}}+L_4\f{d+2}{d-2}\left(\f{1}{\Delta}\right)^{-\f{d}{2}}\bigg].
\end{align}
We should note that using the cut-off regularization, one can easily see that the three terms of the above relation are divergent as $\Lambda^4$, $\Lambda^2$ and $\log \Lambda$, respectively.
However, when using the dimensional regularization, all of these divergences appear in a same form and cannot be distinguished, as we will see in the following.
It is obvious that the factor $\Gamma\left(2-\f{d}{2}\right)$ is singular for ordinary four-dimensional space-time ($d=4$).
The prescription of the dimensional regularization to deal with this problem is to take $d=4-\epsilon$ where $\epsilon$ is a small positive parameter which should approach zero to restore the four-dimensional space-time.
Using this replacement and arranging the terms in different orders of $\epsilon$, we obtain
\begin{align}\label{13}
\mc{X}_3=&\f{i\alpha_3^2}{2(4\pi)^{2}}\int d^4k\int_0^1 dx\bigg[2(L_0-2 L_2\Delta+3 L_4\Delta^2)\f{1}{\epsilon}\nonumber\\
+&L_0(\gamma-\log\Delta)- L_2\Delta(2\gamma-2\log\Delta+1)+ L_4\Delta^2 (3\gamma-3\log\Delta+2)+\mc{O}\left(\epsilon\right)\bigg],
\end{align}
where we have used the following relation
\begin{align}\label{14}
\frac{\Delta^{-\frac{\epsilon}{2}}}{(4\pi)^{2-\frac{\epsilon}{2}}}\Gamma\left(\f{\epsilon}{2}\right)=\frac{1}{(4\pi)^2}\left(\frac{2}{\epsilon}-\log\Delta
-\gamma+\log (4\pi)+\mc{O}(\epsilon)\right),
\end{align}
in which $\gamma$ is the Euler-Mascheroni constant.

In the limit $\epsilon\rightarrow0$, the first term in (\ref{13}) diverges, but the other terms remain finite. Performing the integration over $x$, which is an easy task, the divergent term becomes 
\begin{align}
\mc{X}^{inf}_3=\f{i\alpha_3^2}{2(4\pi)^{2}}&\int d^4k\, \f{2}{\epsilon}\bigg[\left(445+\f{3071}{15}\f{k^2}{m^2}+\f{5527}{210}\f{k^4}{m^4}\right)A^\mu(k)A^\nu(-k)k_\mu k_\nu\nonumber\\
&+\left(\f{256}{3}k^2+76m^2+\f{338}{15}\f{k^4}{m^2}+\f{68}{35}\f{k^6}{m^4}\right)A^\mu(k)A_\mu(-k)\bigg].
\end{align}

Now, to find the divergent part of the second term in \eqref{9}, we do a similar calculation which is described below.
By Fourier transforming the fields and taking the trace in this term, one can obtain
\begin{align}
\hspace{-0.1cm}
\mc{X}_4&\equiv \,\alpha_4\textmd{tr}\left[\mc{O}^{-1\mu\rho}\mc{L}^{cl}_{4\rho\sigma}\right]\nonumber\\
& =-\alpha_4\! \int \!\f{d^4p}{(2\pi)^4}\!\int \!d^4k\, \f{1}{p^2+m^2}\left[\eta^{\mu\nu}+\f{p^\mu p^\nu}{m^2}\right]\!\bigg[\big(p_\mu p_\nu +c_1(p^2+k^2)\eta_{\mu\nu}-(1+c_1)p_\mu p_\nu\big)\eta_{\alpha\beta}\nonumber\\&
+4 p_\nu k_\beta \eta_{\alpha\mu}+4c_1(k.p)\eta_{\alpha \mu}\eta_{\beta\nu} -c_1 k_\alpha k_\beta \eta_{\mu\nu}-4(1+c_1)p_\beta k_\nu \eta_{\alpha\mu}\bigg]A^\alpha (k) A^\beta (-k),
\end{align}
Again, we perform the integration over $p$ by means of the dimensional regularization method. To do so, we set $d=4-\epsilon$ and after using (\ref{14}), with $\Delta=m^2$, we obtain the following expression:
\begin{align}
\mc{X}_4= -\f{3i\alpha_4}{(4\pi)^2}c_1 m^2\int d^4k &\left[\f{2}{\epsilon}+1-\log(m^2)+\gamma-\log(4\pi)+\mc{O}(\epsilon)\right]\nonumber\\
&\times \left[(k^2-m^2)\eta_{\alpha\beta}-k_\alpha k_\beta\right]A^\alpha(k)A^\beta(-k).
\end{align}
All the terms multiplying by $\f{1}{\epsilon}$ are divergent as $\epsilon$ goes to zero. We call these terms $\mc{X}_4^{inf}$.

Finally, putting the results of the above calculations together, we arrive at the one-loop correction to the propagator of the vector field, which can be expressed as  
\begin{align}
\Gamma^{2pt}_{1}=\mc{X}_3+\mc{X}_4,
\end{align}
the finite part of which can be simplified as follows:
\begin{align}\label{15}
\mc{X}_3^{fin}+\mc{X}_4^{fin}=\f{i}{(4 \pi)^2}\int d^4 k \bigg\{ \f{\alpha_3^2}{2}&\left(m^2 N_1\eta_{\alpha\beta}+N_2 k_\alpha k_\beta\right)-3\alpha_4 c_1 m^2 N_3 [(k^2 -m^2) \eta_{\alpha\beta}\nonumber\\
&-k_\alpha k_\beta]\bigg\}A^\alpha(k)A^\beta(-k),
\end{align}
where
\begin{align}
N_1=\sqrt{\f{-4 m^2-k^2}{k^2}}&\left(-83.50 - 89.30 \f{k^2}{m^2} - 28.99 \f{k^4}{m^4} - 2.97 \f{k^6}{m^6}\right)\tan^{-1}\left(\sqrt{\f{-k^2}{4 m^2+k^2}}\right)\nonumber\\
&+ \left(-68 - 74.67 \f{k^2}{m^2} - 17.47 \f{k^4}{m^4} - 1.49 \f{k^6}{m^6}\right) \log\left(m^2\right)\nonumber\\
&+168.75 + 178.69 \f{k^2}{m^2} + 46.28 \f{k^4}{m^4} + 
 4.18 \f{k^6}{m^6},
\end{align}
\begin{align}
N_2=\f{m^4}{k^4}&\sqrt{\f{-k^2}{4 m^2+k^2}}\bigg(243.81 + 1457.14 \f{k^2}{m^2} + 1029.43 \f{k^4}{m^4} + 255.24 \f{k^6}{m^6} + 21.29 \f{k^8}{m^8}\bigg)\nonumber\\
&\times\tan^{-1}\left(\sqrt{\f{-k^2}{4 m^2+k^2}}\right) + \bigg(-348.33 - 70.33 \f{k^2}{m^2} - 10.64 \f{k^4}{m^4}\bigg)\log\left(m^2\right)\nonumber\\
&+\f{m^2}{k^2}\bigg(60.95 + 795.69 \f{k^2}{m^2} + 234.64 \f{k^4}{m^4} + 31.66 \f{k^6}{m^6}\bigg),
\end{align}
\begin{align}
N_3=-0.95-\log\left(m^2\right).
\end{align}
It is important to notice that some of terms in Eq.\,(\ref{15}) have the same shape as the original terms of the Lagrangian. 
This means that the propagator in this problem recieves some corrections and this is in contrast to the result that has been obtained in the case of scalar Galileons.
%*************************************************************
\section{Concluding remarks}\label{conclu}
In this paper, we have added the vector Galileon terms to the Proca theory to study the effective action of a massive vector field with Galileon self
interactions. 
The first vector Galileon term is actually the mass term for the vector field.  
We have calculated the one-loop quantum correction to the two-point function of the vector field. 
The method chosen in this work is the background field method in which one perturbs the field, here the vector field, from its classical value and then find the one-loop effective action. 
It is obvious that only terms with four powers of vector field contribute to the  one-loop correction to the propagator of the theory. 
For this reason we have neglected the last vector Galileon interaction  $\mc{L}_5$ which has five powers of vector field. 
This term would play role in calculations such as the two-loop correction to propagator or corrections to the three-point vertex. 
However, in our problem only $\mc{L}_3^2$ and $\mc{L}_4$ contribute and these terms correspond to these Feynman diagrams:
\begin{figure}[h]
\centering
\includegraphics[scale=0.3]{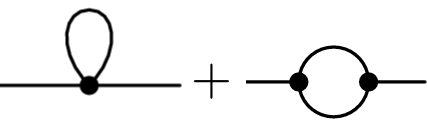}
\end{figure}

To deal with divergent parts of the loop integrals, we have applied the dimensional regularization technique. 
Using the cut-off regularization it is easy to show that there are three kinds of divergences in these terms, the higher of which is $\Lambda^4$, but in the dimensional regularization all of these divergences show up as $1/\epsilon$. 
Doing all these calculations, we have obtained the finite and divergent parts of the integrals. 
The divergent parts define counter-terms that should be subtracted from the effective action in order to have a finite action and the finite parts are added to the effective action. 
Surprisingly, some of these terms are in the shape of the terms in the uncorrected Lagrangian in contrast to the result obtained for the scalar Galileons \cite{nonrenor,prdplb}. 
For Lagrangian with the scalar Galileons, it has been found that none of the correction terms are the same as the old terms in the Lagrangian. 
Consequently, these corrections do not change the propagator and only add some new terms to the Lagrangian. 
This statement is known as the non-renormalization theorem. By this argument, our calculations end up with a very important conclusion: there is no non-renormalization theorem for vector Galileons.

In this work, we have considered a special choice $f_i=\mc{A}^2$. It is worth mentioning that any choice for $f_i$'s which causes that the Lagrangian $\mc{L}_i$ acquires more than 4 powers of $\mc{A}_\mu$ is not interesting, because it will not contribute to our calculations. This situation can happen for any other Lagrangian. If we had assumed that $f_4=1$, then $\mc{L}_4^2$ will contribute the effective action and modify the propagator again.

In fact, in order to investigate the non-renormalization theorem, we should also calculate the corrections to $\alpha_3$, $\alpha_4$ and $\alpha_5$. In this paper, we have seen that the propagator is corrected. So there is no need to calculate corrections to the other terms.
\section*{Acknowledgement}
We would like to thank Gregory Gabadadze for very useful comments.
\appendix
\section*{Appendix}
The  $L^{\alpha\beta}_i$ appearing in equation \eqref{11} are as follows:

\begin{align}
L_{0}^{\alpha\beta}=\Bigg[-\left(\f{k^2}{m^2}+4\right)+&2x+\left(8+10\f{k^2}{m^2}-\f{k^4}{m^4}\right)x^2+20\f{k^2}{m^2} x^3\nonumber\\&+\f{k^2}{m^2}\left(16+11\f{k^2}{m^2}\right)x^4+18\f{k^4}{m^4}x^5+8\f{k^4}{m^4}x^6\Bigg]k^\alpha k^\beta,\nonumber
\end{align}
\begin{align}
L_{2}^{\alpha\beta}&=2\bigg[1+\f{k^2}{m^2}x+2\f{k^2}{m^2}x^2+\f{k^4}{m^4}x^3+\f{k^4}{m^4}x^4\bigg]\eta^{\alpha\beta}\nonumber\\
&+\f{1}{m^2}\Bigg[\f52-\f14 \f{k^2}{m^2}+\left(22-3\f{k^2}{m^2}\right)x+\left(32+\f{45}{2}\f{k^2}{m^2}\right)x^2+64\f{k^2}{m^2} x^3+40 \f{k^2}{m^2} x^4\Bigg]k^\alpha k^\beta,\nonumber
\end{align}
\begin{align}
L_{4}^{\alpha\beta}=\f{1}{m^4}\left[8(4k^2x^2+2k^2x+3m^2)\eta^{\alpha\beta}+(160x^2+152x+3)k^\alpha k^\beta\right].\nonumber
\end{align}

\end{document}